\documentclass{jfm}
\usepackage{graphicx}
\usepackage{epstopdf, epsfig}
\usepackage{amsmath}
\usepackage{graphicx}
\usepackage{lineno}
\usepackage{multirow}
\usepackage{caption}
\usepackage{subcaption}
\usepackage{float}
\usepackage[utf8]{inputenc}
\usepackage{array}
\DeclareGraphicsExtensions{.pdf,.png,.jpg,.eps}
\usepackage{epstopdf}
\usepackage{xcolor}
\usepackage{multimedia}

\shorttitle{Stability of EVP plane Couette flow}
\shortauthor{R. Patne}

  \title{Stability of elastoviscoplastic plane Couette flow}
   \author{Ramkarn Patne\aff{1}\corresp{\email{ramkarn@che.iith.ac.in}}}

 \affiliation{\aff{1}Department of Chemical Engineering, Indian Institute of Technology Hyderabad, Kandi, Sangareddy, Telangana 502285, India}

\begin{document}

\maketitle

\begin{abstract}

Several studies have investigated the turbulent flow of elastoviscoplastic (EVP) fluids, which exhibit yield stress in addition to viscoelasticity. The instabilities that could be responsible for the transition to turbulence in the EVP fluid flows remain unknown. Thus, the present explores the linear stability of EVP plane Couette flow (PCF) by employing Saramito model. The eigenvalue problem is solved by using pseudo-spectral method. In the limit of vanishing yield stress, EVP fluid behaves as Upper Convected Maxwell (UCM) fluid. The creeping flow of UCM fluid exhibits two stable Gorodotsov \& Leonov (GL) modes, thus a stable flow. As the Bingham number (i.e., yield stress) increases, the GL modes become unstable, implying an unstable flow. Additionally, there are new unstable modes with phase speed equalling the average velocity of the fluid. The analysis reveals an extra tangential stress term, arising due to yield stress, is responsible for the predicted instabilities. Also, the Saramito model exhibits weak Hadamard instability, i.e., unstable perturbations of arbitrarily small wavelengths. The present study demonstrates the removal of the Hadamard instability by adding a stress diffusion term in the Saramito constitutive equation. To conclude, the PCF of an EVP fluid is linearly unstable.

\end{abstract}


\section{Introduction}  \label{sec:Introduction}

Elastoviscoplastic (EVP) fluids or viscoelastic fluids possessing yield stress are common in nature (e.g., mud and lava) \citep{Abdelgawad-et-al-2023}, industrial applications (e.g., Carbopol gels and concrete) \citep{Barry-Meyer-1979, Abdelgawad-et-al-2023}  and biosystems (e.g., blood and airway mucus) \citep{Apostolidis-Beris-2014, Erken-et-al-2023, Shemilt-et-al-2022, Shemilt-et-al-2023}. Recently, \cite{Rosti-et-al-2018,Izbassarov-et-al-2021,Abdelgawad-et-al-2023} have investigated the turbulent flow of EVP flows. However, the instabilities which could be responsible for the turbulent flow remain a mystery. The present study is an effort to address this missing gap. Below, we review the relevant literature.

\subsection{Bingham fluid flows}

An EVP fluid exhibits yield stress; thus, below, we review the relevant literature.
\cite{bingham-1922} proposed a
one-dimensional constitutive model to describe the flow of yield stress fluids. His model assumed that the material behaves as a rigid body before
yielding and as a Newtonian fluid after yielding. The stability of plane Poiseuille flow of Bingham fluid was first investigated by \cite{frigaard-et-al-1994}.
Their analysis predicted that the critical Reynolds number for the onset of unstable flow increases with increasing Bingham number. Later,  \cite{nouar-et-al-2007b} studied the same problem, but he employed the Chebyshev collocation method to solve the eigenvalue problem. Their analysis predicted that the plane Poiseuille flow of a Bingham fluid is linearly stable.
The disagreement between the predictions of \cite{frigaard-et-al-1994} and \cite{nouar-et-al-2007b} was due to the erroneous
numerical method employed by the former to compute the eigenvalues. Note that the plane Poiseuille
flow of a Newtonian fluid exhibits Tollmien-Schlichting instability with critical Reynolds number $5772$. This clearly demonstrates the role of yield stress in suppressing instabilities. 

\cite{Peng-Zhu-2004} investigated the linear stability of Bingham fluids in spiral Couette flow and concluded the stabilizing effect of yield stress on flow. \cite{Landry-et-al-2005} investigated the stability of Taylor–Couette flows of a Bingham fluid by considering three cases, viz., fully unyielded, partially yielded and fully yielded. Only solid body rotation is possible in the fully unyielded case since the Bingham fluid treats unyielded fluid as a rigid body. This will arise if the imparted stresses are lower than the yield stress of the fluid through the annular gap. If the imparted stress is sufficient to yield part of the fluid region, then the partially yielded case arises. If the imparted stress exceeds the yield stress of fluid throughout the fluid domain, then the fluid is fully yielded. A similar case of fully yielded Bingham fluid has been previously considered by \cite{Sahu-et-al-2007} while analysing the two-layer flow of Newtonian and Bingham fluids. In the present study, following \cite{Landry-et-al-2005,Sahu-et-al-2007} a fully yielded flow is considered as described in Sec.~\ref{sec:problem-formulation}. An excellent review of studies investigating the stability of Bingham fluid flows is given by \cite{balmforth-et-al-2014}.

\subsection{Viscoelastic fluid flows}

In addition to yield stress, an EVP fluid also exhibits elasticity. Thus, next, we review the relevant literature dealing with the stability of viscoelastic fluid flows. 

\cite{gorodtsov-leonov-1967} analysed the linear stability of Upper Convected Maxwell (UCM) plane Couette flow (PCF) in the creeping-flow, i.e., in inertialess limit. They predicted two stable discrete modes (henceforth referred to as \emph{GL} modes). \cite{wilson-et-al-1999} extended analysis of \cite{gorodtsov-leonov-1967} to include solvent and predicted a stable flow for $W \sim O(1)$.
\cite{renardy86} extended previous analyses to a finite Reynolds number and predicted a linearly stable flow for arbitrary $Re$. \cite{nouar-et-al-2007} analysed the stability of PCF of an inelastic shear-thinning (power-law) fluid and predicted a linearly stable flow at low $Re$.
\cite{grillet-et-al-2002} analysed the PCF of \cite{giesekus-1982} and  \cite{thien-tanner-1977} (PTT) models in the creeping-flow limit. Their analysis predicted that the PCF of a Giesekus fluid is linearly stable. In contradistinction, the PCF of the PTT fluid was found to be linearly unstable for a highly shear-thinning viscoelastic fluid at a high Weissenberg number. 
\cite{arora-et-al-2004} predicted a linearly stable PCF of pom-pom fluid in the creeping-flow limit.

The analysis by \cite{morozov_saarloos2005} predicted a purely elastic subcritical instability in PCF of UCM fluid.  \cite{hoda_jovanovic_kumar_2008,hoda_jovanovic_kumar_2009} explored the transient growth of perturbations for inertialess viscoelastic PCF and suggested a nonmodal route to transition.
\cite{chokshi-kumaran-2009,cromer-et-al-2013,cromer-et-al-2014,Beneitez-et-al-2023,Couchman-et-al-2024}
considered coupling between the polymer stress and concentration and predicted a linearly unstable PCF. This coupling modifies the governing equations by the addition of the polymer concentration convection-diffusion equation. Such a coupling will not be considered in the present study. Recently, \cite{piyush-et-al-2018, Chaudhary-et-al-2019, Chaudhary-et-al-2021, Khalid-et-al-2021, Dong-Zhang-2022} have extensively analysed the impact of fluid elasticity on pressure-driven flows. A detailed literature review of studies concerned with the effect of elasticity on flow dynamics can be found in \cite{Datta-et-al-2022}. \cite{samanta-et-al-2013,choueiri-et-al-2018,chandra-et-al-2018} have experimentally demonstrated the impact of fluid elasticity on flow dynamics.

The combined effect of elasticity and power-law type shear-thinning in the absence of yield stress was analysed by \cite{wilson-rallison-1999, wilson-loridan-2015, castillo-wilson-2017}. \cite{wilson-rallison-1999, wilson-loridan-2015, castillo-wilson-2017} analysed the linear stability of pressure-driven channel flow of \cite{white-metzner-1963} fluid using the power-law model for the viscosity and relaxation time and a constant relaxation modulus. Their analysis predicted an unstable flow provided the power-law index $n$ is less than $0.3$ in the creeping-flow limit. The predicted instability was experimentally observed by \cite{bodiguel-et-al-2015}. The experiments by \cite{poole-2016,picaut-et-al-2017, wen-et-al-2017, chandra-et-al-2019} further demonstrated the existence of shear-thinning elastic instabilities for the flows of concentrated polymer solutions for low Reynolds number.

\subsection{Elastoviscoplastic (EVP) fluid flows}

The fluids encountered in natural settings and industrial processes could possess yield stress and behave as viscoelastic fluid upon yielding. The above-discussed studies consider these properties separately. \cite{Moyers-et-al-2011} analysed the linear stability of Plane Poiseuille flow of an EVP fluid using a revised version of the \cite{Putz-Burghelea-2009} model. Their analysis concluded that stability results are in close conformity with those of a pseudo-plastic fluid. To describe the flows of EVP fluids, \cite{Saramito-2007} developed a constitutive equation which satisfies the second law of thermodynamics. As per the model, the unyielded state of the material is a neo-Hookean solid, and the yielded state is a viscoelastic Oldroyd-B fluid. The \cite{Saramito-2007} model predicts a smooth transition between the solid and liquid state due to yielding transition being based on the von Mises criterion. The predictions of the \cite{Saramito-2007} model are in excellent agreement with the experimental observations \citep{Fraggedakis-et-al-2016}. The \cite{Saramito-2007} model has been extensively utilised to analyse turbulent flow in EVP fluid flows by \cite{Rosti-et-al-2018, Izbassarov-et-al-2021, Abdelgawad-et-al-2023}. To the best of the author's knowledge, the linear stability of EVP fluid modelled by \cite{Saramito-2007} constitutive equation has not been analysed. Owing to this, in the present study, we will utilise the \cite{Saramito-2007} model to analyse the linear stability of EVP plane Couette flow (PCF).

The rest of the paper is organised as follows. The base state quantities and linearised perturbation equations are derived in Sec.~\ref{sec:problem-formulation}. The numerical methodology utilised to solve the eigenvalue problem and its validation is discussed in Sec.~\ref{sec:Numerical methodology}.  Section~\ref {sec:Results and Discussion} discusses the stability results, the physical mechanism responsible for the existence of the predicted instabilities, and the removal of the Hadamard instability. The salient results and implications of the present work are summarised in Sec.~\ref{sec:Conclusions}.

\section{Problem formulation} \label{sec:problem-formulation}

Consider an incompressible EVP fluid of yield stress $\tau_0^*$, viscosity $\eta^*$ and relaxation modulus $G^*$ flowing through the gap of width $R^*$ between two plates. The lower plate (at $y^*=0$) is stationary while the upper plate  (at $y^*=R^*$) is moving at a steady and constant speed $V^*$ in the positive $x$ direction. Here, superscript $`*'$ signifies a dimensional quantity. The dimensional \cite{Saramito-2007} model is
\begin{eqnarray}
\nonumber
\frac{1}{G^*} \left[ \frac{\partial \boldsymbol{\tau}^*}{\partial t^*} + (\mathbf{v}^* \cdot{\nabla^*}) \boldsymbol{\tau}^* - (\nabla^* \mathbf{v^*})^{T} \cdot\boldsymbol{\tau}^*-\boldsymbol{\tau}^* \cdot (\nabla^* \mathbf{v}^*) \right] +\\
\max\left(0, \left[\frac{|\boldsymbol{\tau}_d^*|-\tau_0^*}{\eta^*|\boldsymbol{\tau}_d^*|} \right] \right)\boldsymbol{\tau}^*=  \boldsymbol{\dot \gamma}^*, \label{eq:saramito-equation-dimensional}
\end{eqnarray}
\noindent
where $\boldsymbol{\tau}^*$ is the stress field. The deviatoric stress tensor $\boldsymbol{\tau}_d^* = \boldsymbol{\tau}^* - 1/N tr(\boldsymbol{\tau}^*)$ where $N$ is the number of dimensions and $|\boldsymbol{\tau}_d^*| = \sqrt{1/2 \, \boldsymbol{\tau}_d^* : \boldsymbol{\tau}_d^*}$ \citep{Saramito-2007,Fraggedakis-et-al-2016}. The strain rate tensor is $\boldsymbol{\dot \gamma}^*=(\nabla^*\mathbf{v}^*)+(\nabla^*\mathbf{v}^*)^{T}$. As per the von Mises criterion followed by \cite{Saramito-2007} model, when $\boldsymbol{\tau}_d^*$ exceeds the yield stress of the fluid, the material starts flowing obeying the Maxwell constitutive equation.

\begin{subequations}
We scale the lengths and velocities by $R^*, V^*$ while the stresses and pressure are scaled by $\eta^* V^*/R^*$. Let $\mathbf{v}=(v_x,v_y,v_z)$ be the velocity field, and $p$ is the pressure field in the fluid. For the sake of simplicity, we assume the absence of solvent and creeping-flow limit.
Thus, the dimensionless governing equations are
\begin{eqnarray}
\nabla \cdot \mathbf{v}=0,~~~ \label{eq:continuity-equation}\\
 -\nabla p + \nabla \cdot \boldsymbol{\tau }=0,~~~ \\
W \left[ \frac{\partial \boldsymbol{\tau}}{\partial t} + (\mathbf{v} \cdot{\nabla}) \boldsymbol{\tau}- (\nabla\mathbf{v})^T \cdot\boldsymbol{\tau}-\boldsymbol{\tau} \cdot (\nabla\mathbf{v}) \right] + \max \left(0, \left[\frac{|\boldsymbol{\tau}_d|-B}{|\boldsymbol{\tau}_d|} \right] \right) \boldsymbol{\tau}  = \boldsymbol{\dot \gamma},~~~ \label{eq:SM-equation}
\end{eqnarray}
\noindent
where, $B = \tau_0^* R/(\eta^* V^*)$ is the Bingham number, $W = \eta^* V^*/(G^*R^*)$ is the Weissenberg number and $\nabla$ is the gradient operator. 
\label{eq:goveq}
\end{subequations}

\begin{subequations}

The base state quantities for the flow under consideration have been obtained by \cite{Fraggedakis-et-al-2016} using the following procedure.
For PCF, owing to the constant shear stress throughout the fluid domain, either the fluid can be fully unyielded or yielded depending on the value of $|\boldsymbol{\tau}_d^*| $ and $\tau_0^*$. In the present study, we consider fully yielded fluid for which $|\boldsymbol{\tau}_d^*| > \tau_0^*$. Note that previously, \cite{Landry-et-al-2005} considered a similar fully yielded flow for Taylor-Couette flow while \cite{Sahu-et-al-2007} considered a fully yielded flow for a two-layer flow of Bingham and Newtonian fluids. For a fully developed, two-dimensional steady flow, the base state quantities are
\begin{eqnarray}
\bar v_x=  y; \quad
\bar \tau_{xx}= 2 W \bar \tau_{xy}^2; \quad \bar \tau_{yy}=0,\\
  {\cal F} \, \bar \tau_{xy}=1; \quad {\cal F} =\frac{|\boldsymbol{\bar \tau_d}|-B}{|\boldsymbol{\bar \tau_d}|}; \quad |\boldsymbol{\bar \tau_d}| = \sqrt{\frac{\bar \tau_{xx}^2}{4} + \bar \tau_{xy}^2}.
\end{eqnarray}
\noindent 
To obtain base state stress tensor components, the above equation must be solved for $\bar \tau_{xy}$ for specified values of $W, B$ and $n$ using a root-finding algorithm. In the present study, we employ \emph{fsolve} MATLAB function to solve the above transcendental set of equations for $\bar \tau_{xy}$. The obtained root is further confirmed by \emph{FindRoot} and \emph{NSolve} functions in MATHEMATICA. For $B=0$, the above equations yield $\bar \tau_{xy}=1$ and $\bar \tau_{xx}=2W$ applicable to PCF of Upper Convected Maxwell fluid. \cite{Fraggedakis-et-al-2016} utilised Newton's method to solve the above transcendental equations.
    \label{eq:base-state}
\end{subequations}

For the sake of simplicity, we assume two-dimensional perturbations, $f'(\mathbf{x},t)= \tilde f(y) \, e^{i\, k \,(x-ct)}$ imposed on the above base-state where $f'(\mathbf{x},t)$ is any perturbation to dynamical variable and $\tilde f(y)$ is the corresponding eigenfunction. Here, $k$ and $c=c_r+i c_i$ are the wavenumber and complex wavespeed, respectively. The flow is unstable if at least one eigenvalue satisfies the condition $c_i>0$.

\begin{subequations}
The governing equations \eqref{eq:goveq} are linearised around the base state \eqref{eq:base-state}. In the resulting linearised equations, the normal modes are substituted to obtain
\begin{eqnarray}
ik \tilde v_x+D\tilde v_y=0, \label{eq:ppf-ConEq}\\
-ik\tilde p+ik\tilde \tau_{xx}+D\tilde \tau_{xy}=0, \label{eq:ppf-x-mom}\\
-D\tilde p+ik\tilde \tau_{xy}+D\tilde \tau_{yy}=0, \label{eq:ppf-y-mom}
\end{eqnarray}
\noindent
where $D=d/dy$. The constitutive equation~(\ref{eq:SM-equation}) gives
\begin{eqnarray}
\nonumber
W [ik(\bar v_x-c) \tilde \tau_{xx}+\tilde v_y D\bar \tau_{xx}-2ik \bar \tau_{xx} \tilde v_x-2 D\bar v_x \tilde \tau_{xy}-2 \bar \tau_{xy} D \tilde v_x]\\
+ {\cal F } [1+(t_4-t_2) \bar \tau_{xx}]\tilde \tau_{xx} + {\cal F }(t_3-t_1) \bar \tau_{xx}\tilde \tau_{xy} - {\cal F }(t_4-t_2) \bar \tau_{xx}\tilde \tau_{yy}=2ik\tilde v_x, \label{eq:pf-Txx}\\
\nonumber
W [ik(\bar v_x-c) \tilde \tau_{xy}+\tilde v_y D\bar \tau_{xy} -ik \bar \tau_{xy} \tilde v_x-D\bar v_x \tilde \tau_{yy}-ik \bar \tau_{xx} \tilde v_y-\bar \tau_{xy} D\tilde v_y]\\
+ {\cal F } [1+(t_3-t_1) \bar \tau_{xy}]\tilde \tau_{xy} + {\cal F } \bar \tau_{xy}(t_4-t_2) (\tilde \tau_{xx}-\tilde \tau_{yy})=  (ik\tilde v_y+D\tilde v_x), \label{eq:pf-Txy}\\
W [ik(\bar v_x-c) \tilde \tau_{yy}-2ik \bar \tau_{xy} \tilde v_y]+ {\cal F }\tilde \tau_{yy}=2D\tilde v_y , \label{eq:pf-Tyy}
\end{eqnarray}
\noindent
 where
 \begin{eqnarray}
     t_1=\frac{\bar \tau_{xy}}{|\boldsymbol{\bar \tau_d}|^2 }; \quad t_2=\frac{\bar \tau_{xx}}{4 |\boldsymbol{\bar \tau_d}|^2 };\quad
    t_3=\frac{\bar \tau_{xy}}{|\boldsymbol{\bar \tau_d}| (|\boldsymbol{\bar \tau_d}|-B) }; \quad t_4=\frac{\bar \tau_{xx}}{4 |\boldsymbol{\bar \tau_d}| (|\boldsymbol{\bar \tau_d}|-B)}, \label{ts}
 \end{eqnarray}
 are the parameters arising during linearisation. The above equations are subjected to the following boundary conditions. For $B=0$, from the above expressions, $t_1=t_3$ and $t_2=t_4$ which will remove the additional terms in the linearised constitutive equations \eqref{eq:pf-Txx}-\eqref{eq:pf-Tyy} and reduce ${\cal F }=1$. Thus, the additional terms in equations \eqref{eq:pf-Txx}-\eqref{eq:pf-Tyy} compared to a UCM fluid are clearly due to nonzero yield stress. 
\label{eq:linearised-PCF}
\end{subequations}

 \begin{subequations}
 At both plates ($y=0,1$), we impose no-slip and impermeability conditions, implying
 \begin{eqnarray}
     \tilde v_x = 0; \quad \tilde v_y = 0. \label{eq:bc-PCF}
 \end{eqnarray}
      \label{bcs}
 \end{subequations}

\section{Numerical methodology} \label{sec:Numerical methodology}

In the pseudo-spectral method, the eigenfunctions for each dynamic field are expanded into a series of the Chebyshev polynomials as 
\begin{eqnarray}
 \widetilde{f}(y)=\sum_{n=0}^{n=N} b_n {\cal T}_n (y),
\end{eqnarray}
where ${\cal T}_n(y)$ are Chebyshev polynomials of degree $n$, $b_n$ are series coefficients, and $N$ is the highest degree of the polynomial in the series expansion or, equivalently, the number of collocation points. Upon substitution of the above series expansion in \eqref{eq:linearised-PCF} and \eqref{bcs} results in an eigenvalue problem of the form
\begin{eqnarray}
\mathbf{A}\mathbf{e} + c \mathbf{B}\mathbf{e} =0,
\label{geneig}
\end{eqnarray}
where $\mathbf{A}$ and $\mathbf{B}$ are the discretised matrices  
and $\mathbf{e}$ is the vector of $b_n$. 

The details of the discretisation procedure and construction of the matrices can be found in \cite{trefethen-2000} and \cite{schmid-henningson-2001}. 
We use the \emph{eig} MATLAB routine to solve the eigenvalue problem~(\ref{geneig}). 
To filter out the spurious modes from the genuine, numerically computed spectrum of the problem, the latter is determined for $N$ and $N+2$ collocation points, and the eigenvalues are compared with a priori specified tolerance $10^{-5}$. 

As mentioned in Sec.~\ref{sec:Introduction}, the PCF of a UCM fluid exhibits two discrete, stable GL modes in the creeping-flow limit. Following \cite{gorodtsov-leonov-1967}, these modes can be analytically calculated as follows. To remove the yield stress effect, substitute $B=0$ in equations \eqref{eq:linearised-PCF}. After some algebraic manipulation the resulting equations reduce to \citep{gorodtsov-leonov-1967}
\begin{eqnarray}
(y_n^2 D^2 - 2y_n D-k^2y_n^2+2)(D^2+2ikD-2k^2W^2-k^2)\widetilde v_y =0,
\end{eqnarray}
where $y_n=y-c-i/(kW)$. The solution of the above differential equation is 
\begin{eqnarray}
\nonumber
\widetilde v_y = \frac{ a_1(c-y_n)}{2 k^2 W (i + W)} \exp\left(k c + i/W - k y_n \right) 
+ \frac{a_2 (c - y_n)}{4 k^3 W (-i + W)} \exp\left(-kc - i/W + k y_n \right)  \\
\nonumber
+ a_3 \exp\left[\frac{-\left(i k W + \sqrt{k^2 (1 + W^2)}\right) (-i - kc W + k W y_n )}{k W} \right] \quad\\
+ a_4 \exp\left[\frac{\left(-i k W + \sqrt{k^2 (1 + W^2)}\right) (-i - kc W + k W y_n )}{k W} \right].\quad \label{eq:vy-solution}
\end{eqnarray}
where $a_1,a_2,a_3$ and $a_4$ are integration constants. Substituting the above solution in the boundary conditions \eqref{bcs} leads to a dispersion relation, which can then be solved for the eigenvalue $c$ for specified values of $k$ and $W$ by employing a transcendental root finding scheme in MATLAB.

\begin{figure}
    \centerline{\includegraphics[width=0.65\textwidth]{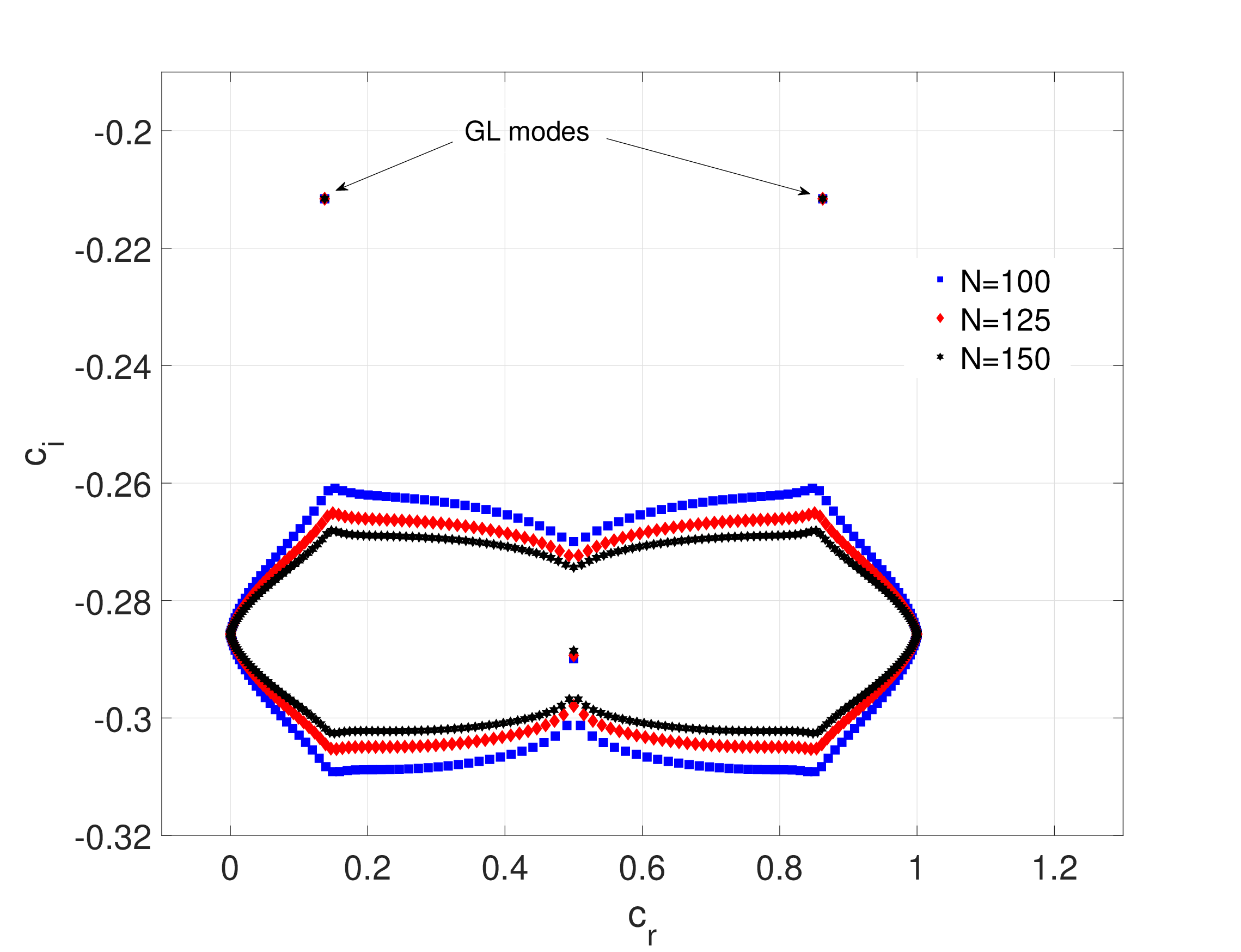}}
    \caption{\small   The variation of the discrete GL modes and continuous spectrum due to variation in the number of collocation points at $W=5, B=0$ and $k=0.7$. The GL modes exhibit negligible variation despite an increase in the number of collocation points.  }
    \label{fig:ci_vs_cr_w5_k0p7_B0}
\end{figure}

From the above discussion, for $B=0$, the present numerical methodology should predict two stable converged eigenvalues as the number of collocation points is varied. This procedure has been implemented in figure~\ref{fig:ci_vs_cr_w5_k0p7_B0} for a set of parameters. Despite a change in the number of collocation points, the discrete GL modes exhibit negligible change. Furthermore, by utilizing the analytical expressions \eqref{eq:vy-solution}, the obtained eigenvalues are in excellent agreement with the numerically predicted eigenvalues, thereby validating the present numerical methodology. On the other hand, the continuous spectrum varies with the change in collocation points. Note that as the Weissenberg number $W$, Bingham number $B$ and wavenumber $k$ increase, the number of collocation points required to capture the most unstable eigenvalue also increases.

\section{Results and Discussion}  \label{sec:Results and Discussion}

\begin{figure}
    \centering
    \begin{subfigure}[b]{0.55\textwidth}
        \centering
        \hspace* {-1.2cm}
        \includegraphics[width=\textwidth]{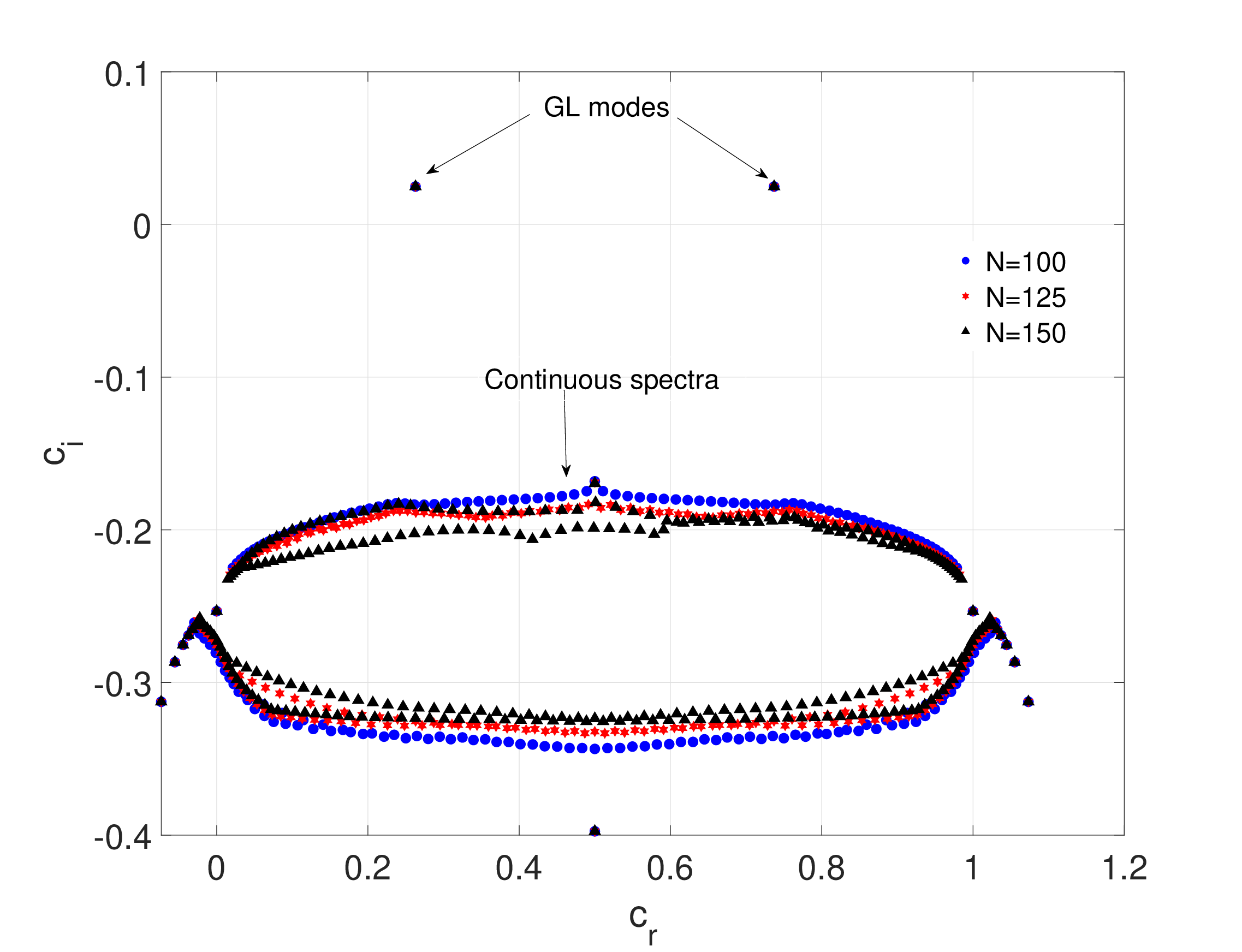}
        \caption{Unfiltered spectra}
        \label{fig:ci_vs_cr_w1_k1_B12}
    \end{subfigure}%
    ~
    \begin{subfigure}[b]{0.55\textwidth}
        \centering
        \hspace* {-1cm}
        \includegraphics[width=\textwidth]{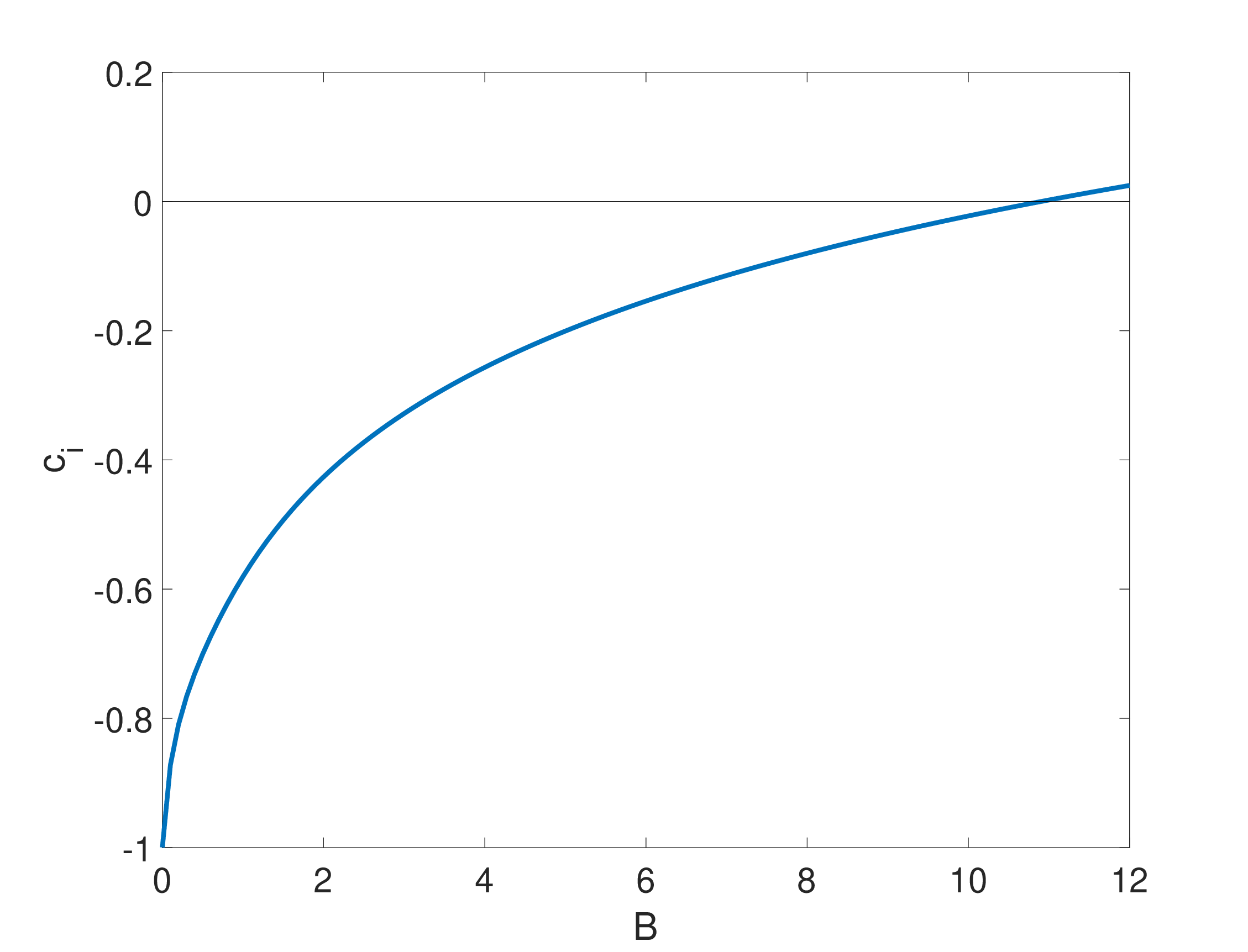} 
        \caption{Variation in $B$}
        \label{fig:ci_vs_B_w1_k1}
    \end{subfigure}
    \caption{ Eigenspectra for $ W=1$ and $k=1$. Panel (a) shows the converged unstable GL modes for $B=12$. Panel (b) demonstrates GL mode destabilisation due to the increasing yield stress effect, i.e., Bingham number $B$. Note that $c_i>0$ implies an unstable mode.}
    \label{fig:spectra GL modes}
\end{figure}

As discussed in Sec.~\ref{sec:Numerical methodology}, a PCF of UCM fluid exhibits two stable GL modes. Thus, first, we analyse the effect of yield stress on the GL modes by varying the Bingham number ($B$) in figure~\ref{fig:spectra GL modes}. From figure~\ref{fig:ci_vs_cr_w1_k1_B12}, for $W=1$ and $B=12$, the GL modes exhibit $c_i>0$, implying unstable GL modes. Further, to confirm the genuine nature of the modes, the spectrum is obtained for three values of the number of collocation points. The convergence of the unstable mode implies the existence of genuinely unstable modes. Further, similar to the stable GL modes, the unstable GL modes also exhibit the same $c_i$ value, albeit with $c_i>0$.

How are we sure that these unstable modes are GL modes? To verify this, the Bingham number is increased from $B=0$ to $B=12$ in steps of $0.1$ as shown in figure~\ref{fig:ci_vs_B_w1_k1}. As $B$ increases beyond $11$, the GL modes become unstable. Note that the same curve is applicable for both GL modes due to the equal $c_i$ value. This clearly establishes that owing to the yield stress, the GL modes become unstable, leading to an unstable flow.

\begin{figure}
    \centerline{\includegraphics[width=0.65\textwidth]{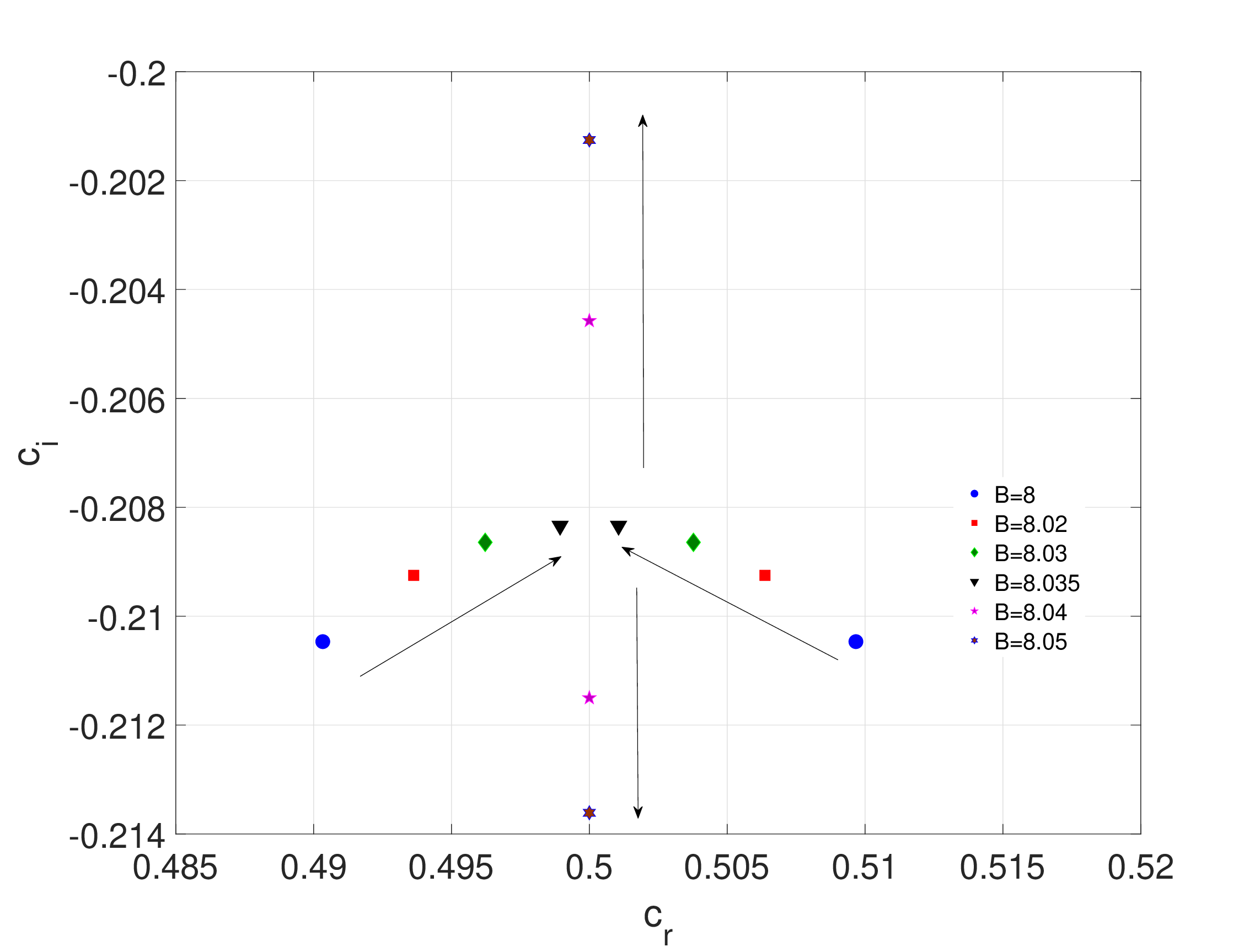}}
    \caption{\small   The manifestation of centre modes due to an increase in the Bingham number for $W=1$ and $k=0.5$. The GL modes merge and give rise to two modes with $c_r=0.5$, henceforth referred to as centre modes. }
    \label{fig:ci_vs_cr_w1_k0p5_B_vary}
\end{figure}

From figure~\ref{fig:ci_vs_cr_w1_k0p5_B_vary}, as the Bingham number is increased,  the $c_r$ values of both GL modes approach $0.5$ for $k<1$. Upon further increase in $B$, the GL modes merge to give rise to two modes with $c_r=0.5$ and different $c_i$ values. The average velocity of the PCF under consideration is $0.5$. The fluid at $y=0.5$ travels at a speed $ =0.5$. Since these new modes travel at an average speed of the fluid, these modes will be referred to as `centre modes' to differentiate them from the GL modes.

\begin{figure}
    \centering
    \begin{subfigure}[b]{0.55\textwidth}
        \centering
        \hspace* {-1.2cm}
        \includegraphics[width=\textwidth]{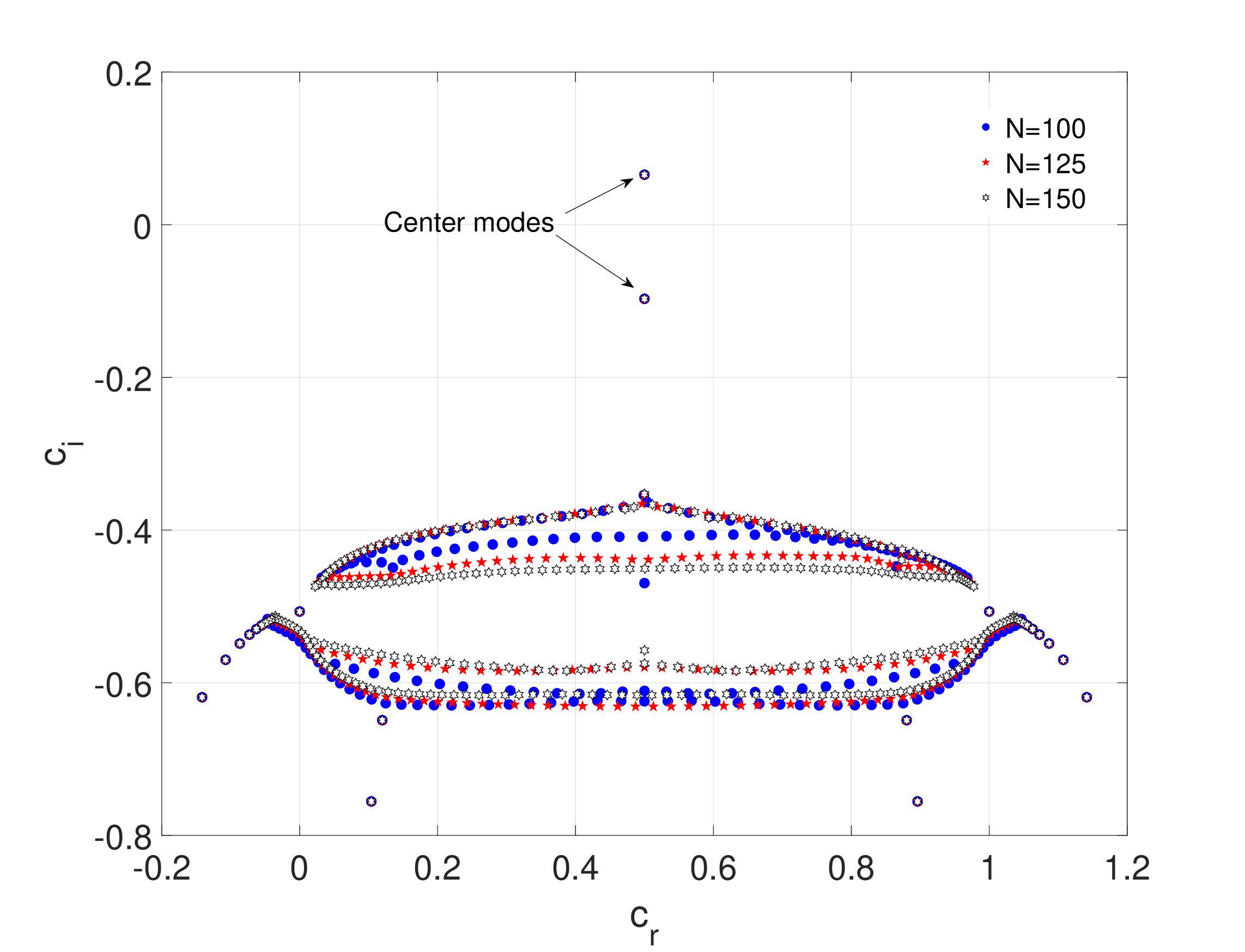}
        \caption{Unfiltered spectra}
        \label{fig:ci_vs_cr_w1_k0p5_B12}
    \end{subfigure}%
    ~
    \begin{subfigure}[b]{0.55\textwidth}
        \centering
        \hspace* {-1cm}
        \includegraphics[width=\textwidth]{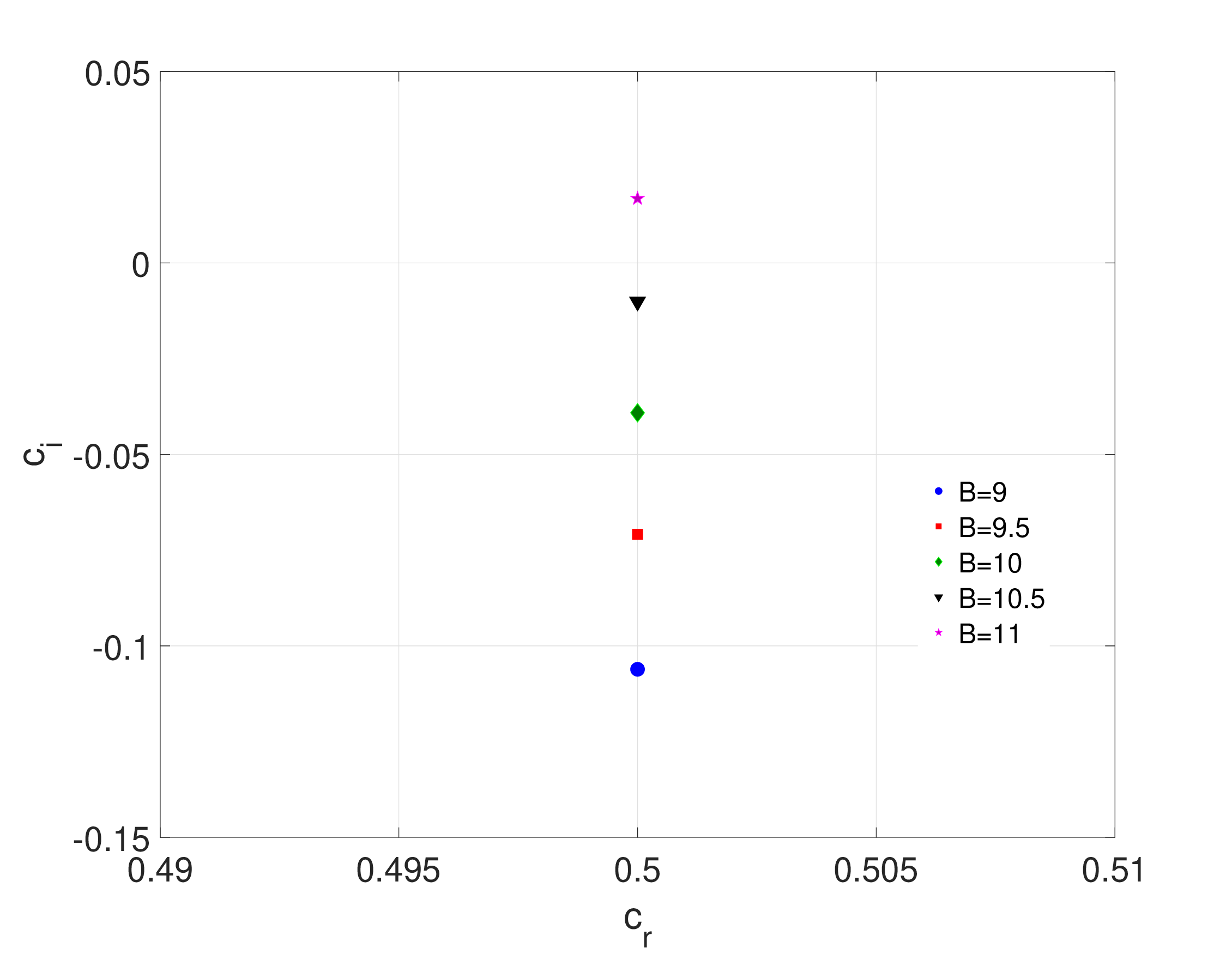} 
        \caption{Variation in $B$}
        \label{fig:ci_vs_cr_w1_k0p5_B_vary_unstable}
    \end{subfigure}
    \caption{ Eigenspectra for $ W=1$ and $k=0.5$ for center modes. Panel (a) shows the converged unstable centre mode for $B=12$. Panel (b) demonstrates centre mode destabilisation due to the increasing Bingham number $B$. Note that $c_i>0$ implies an unstable mode.}
    \label{fig:spectra center modes}
\end{figure}

The new centre modes also become unstable for the same parameters as the GL modes. This aspect is demonstrated in figure~\ref{fig:spectra center modes}. To further ascertain the genuine nature of the centre modes, spectra are obtained for three values of the number of collocation points as shown in figure~\ref{fig:ci_vs_cr_w1_k0p5_B12}. The convergence of the unstable centre mode ascertains the genuine nature. The destabilisation of the centre modes due to an increasing Bingham number is illustrated in figure~\ref{fig:ci_vs_cr_w1_k0p5_B_vary_unstable}. The centre modes become unstable for $B>10.5$, similar to the GL modes. To conclude, along with the GL modes, the yield stress of the fluid also destabilises a new class of centre modes.

\begin{figure}
    \centering
    \begin{subfigure}[b]{0.55\textwidth}
        \centering
        \hspace* {-1.2cm}
        \includegraphics[width=\textwidth]{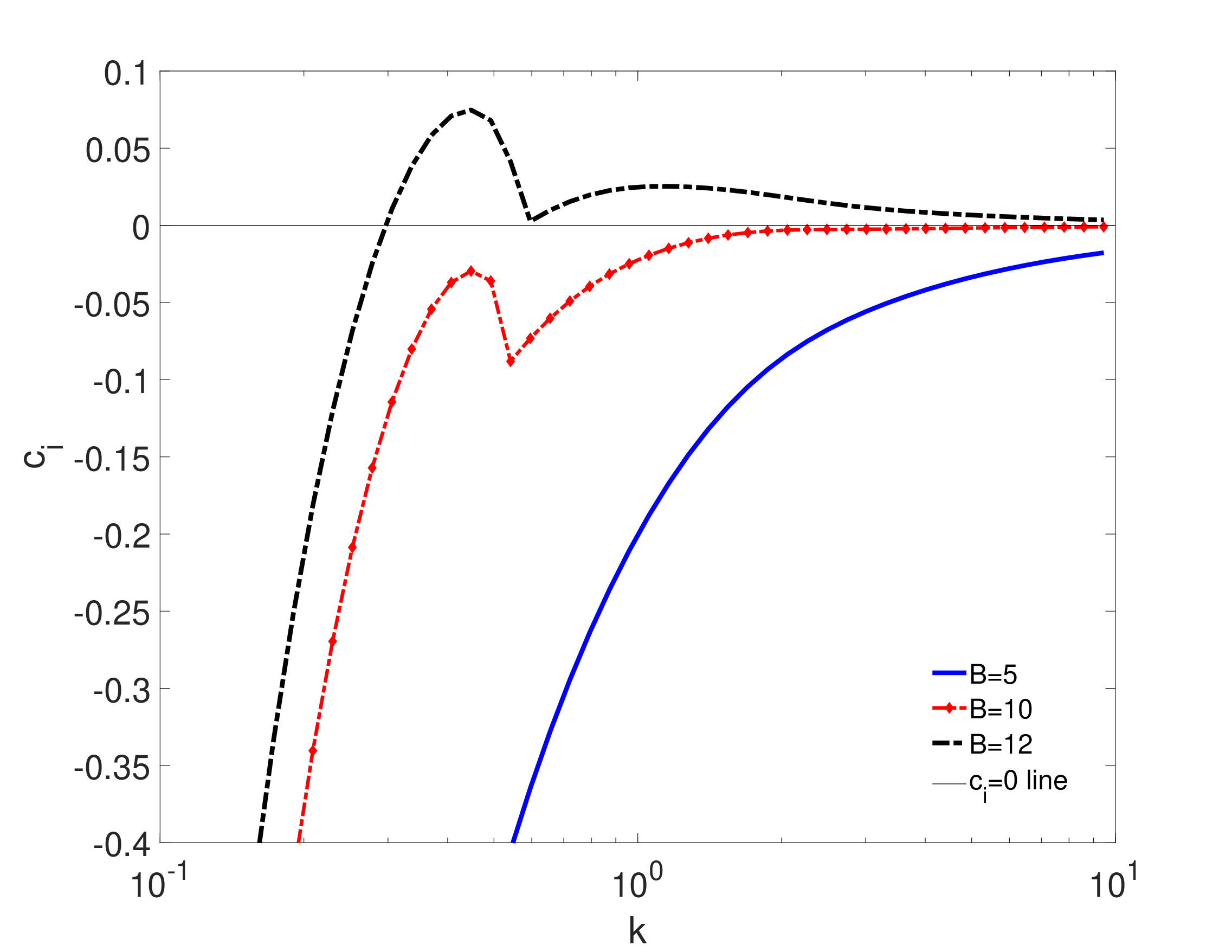}
        \caption{$c_i$ vs $k$}
        \label{fig:ci_vs_k_w1_B_vary}
    \end{subfigure}%
    ~
    \begin{subfigure}[b]{0.55\textwidth}
        \centering
        \hspace* {-1cm}
        \includegraphics[width=\textwidth]{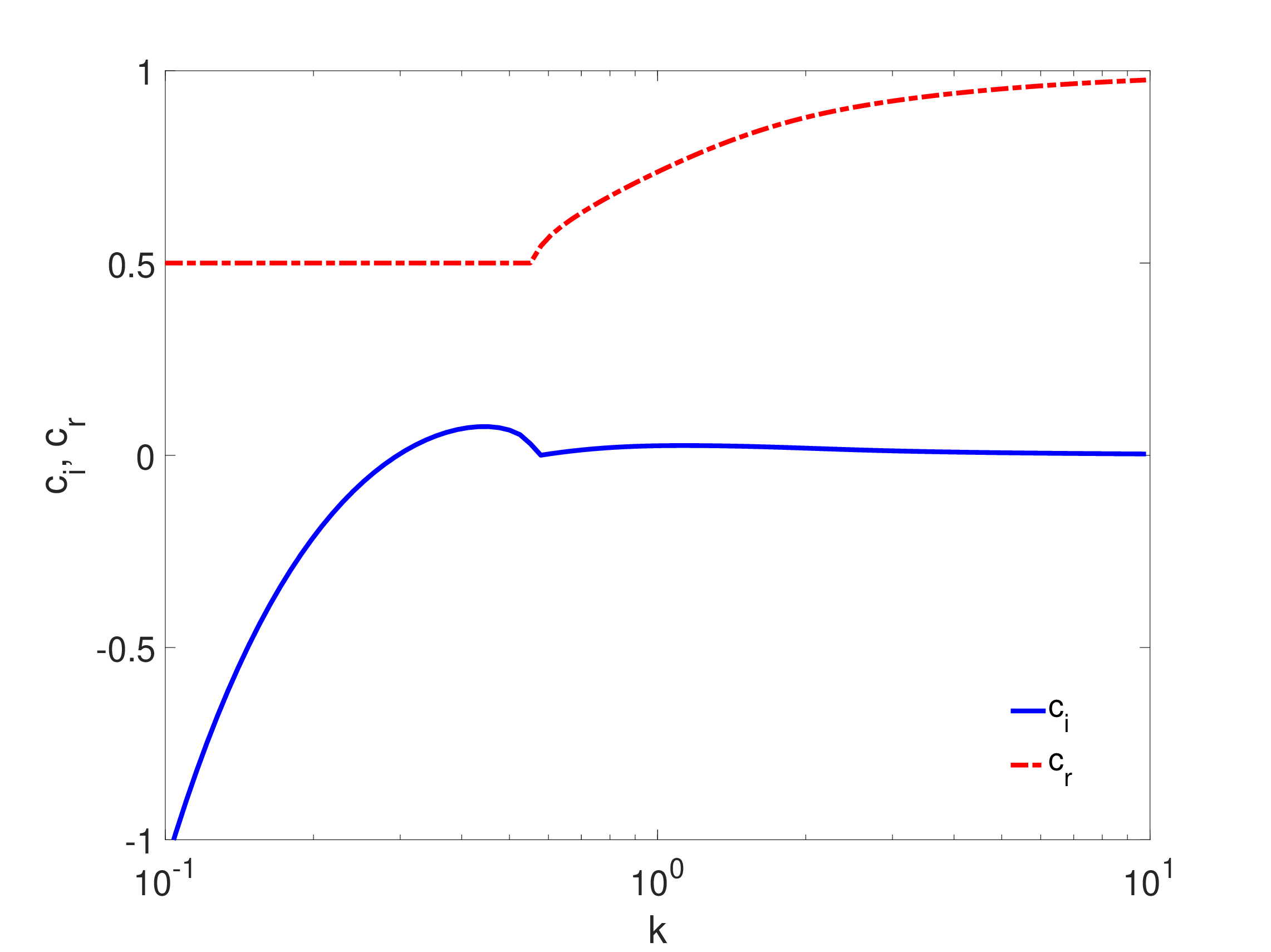} 
        \caption{Mode transition for $B=12$}
        \label{fig:ci_cr_vs_k_w1_B12}
    \end{subfigure}
    \caption{ The dispersion curves for $W=1$. Panel (a) shows the destabilisation of a range of wavenumber as $B$ increases. Panel (b) shows the switching of the most unstable mode from centre mode to GL mode characterised by a change in $c_r=0.5$ to $c_r>0.5$. There is one more GL mode (not shown here) with the same $c_i$ but $c_r<0.5$.}
    \label{fig:dispersion plots}
\end{figure}

The preceding discussion considers the movement of eigenvalues for individual wavenumber as the Bingham number is varied. To obtain a holistic picture for a range of wavenumbers, dispersion plots are shown in figure~\ref{fig:ci_vs_k_w1_B_vary}. For low Bingham numbers, the flow is stable since $c_i<0$ for the whole range of wavenumbers. As the Bingham number is increased, the dispersion curve crosses the $c_i=0$ threshold for a range of wavenumbers, thereby indicating an unstable flow. Additionally, the dispersion curves exhibit two distinct maxima due to the unstable centre and GL modes, i.e., two different types of modes of instability. 

To determine the range of unstable wavenumber due to both types of modes, the variation of both $c_r$ and $c_i$ is plotted in figure~\ref{fig:ci_cr_vs_k_w1_B12}. For $k<0.6$, the mode with $c_r=0.5$ exhibits $c_i>0$, indicating the presence of centre mode. In contrast, for $k>0.6$, the mode with $c_r>0.5$ exhibits $c_i>0$ implying presence of unstable GL mode with $c_r>0.5$. Note that for $k>0.6$, there is one more GL mode that exists with the same $c_i$ but with $c_r<0.5$.

An interesting observation from figure~\ref{fig:ci_vs_k_w1_B_vary} is the existence of instability at arbitrarily high wavenumber with a decreasing $c_i$, a phenomenon referred to as the `weak Hadamard instability' \citep{joseph-saut-1986}. The `Hadamard instability' is generally a consequence of neglecting a dissipative or stabilizing effect such as solvent viscosity \citep{joseph-saut-1986}, interfacial tension \citep{patne-shankar-2018}, or stress diffusivity \citep{patne-et-al-2024}. In the present case, the Hadamard instability can be removed by including stress diffusivity as discussed in Sec.~\ref{sec:Hadamard instability}.

\subsection{Physical mechanism} \label{sec:Physical mechanism}

As mentioned several times in the preceding discussion, the GL modes are linearly stable for a UCM fluid. However, as demonstrated in figures~\ref{fig:ci_vs_B_w1_k1} and \ref{fig:ci_vs_cr_w1_k0p5_B_vary_unstable}, an increasing Bingham number leads to the destabilisation of the GL modes. Also, it gives rise to a new type of centre mode instability. For the existence of an instability, a term containing base state and perturbation quantities is typically responsible. Clearly, the terms which existed for a UCM fluid could not be responsible for the predicted instability owing to the stable GL modes. However, new terms arising as a result of the yield stress in the \cite{Saramito-2007} model could be responsible for this.

To check the term responsible for the predicted instability, the individual term is removed, and its effect on the eigenvalue is observed. This procedure is repeated for all the new terms arising due to the yield stress in equations \eqref{eq:pf-Txx}-\eqref{eq:pf-Tyy}. The exercise concludes that the term ${\cal F } (t_3-t_1) \bar \tau_{xy} \tilde \tau_{xx}$ in equation~(\ref{eq:pf-Txy}) is responsible for the destabilisation. Upon using the base state equations \eqref{eq:base-state}, the term further simplifies to $ (t_3-t_1) \tilde \tau_{xx}$. Thus, this term gives rise to an \emph{extra tangential stress} which leads to the destabilisation. The role of this term in destabilisation is illustrated in table~\ref{table:physical mechanism}. To conclude, the fluid yield stress leads to an extra tangential stress term, viz., $ (t_3-t_1) \tilde \tau_{xx}$ which not only destabilises GL modes but also gives rise to new unstable centre modes. Note that for $B=0$, $t_1=t_3$ and the destabilising term will vanish thereby indicating the role of fluid yield stress.

 \begin{table}
  \begin{center}
    \def~{\hphantom{0}}
    \begin{tabular}{lccc}
        Parameters & With extra tangential stress term & Without extra tangential stress term\\[3pt]
     $W=1,B=12,k=1$ & $0.737074 +    0.02482258 i$ & $0.916623 -     0.140563 i$ \\
     $W=1,B=12,k=0.5$  & $0.500000 +    0.0655863 i$ & $0.833158 -     0.290123 i$ \\
     $W=2,B=20,k=2$ & $0.929200 +    0.01916997 i$ & $0.976722 -     0.0377764 i$ \\
     $W=2,B=20,k=0.3$  & $0.500000 +    0.0347959 i$ & $0.847715 -     0.268481 i$ \\
    \end{tabular}
    \caption{\small A comparison of the most unstable (or least stable) eigenvalue with/without the extra tangential stress term, i.e., $ (t_3-t_1) \tilde \tau_{xx}$ in equation~(\ref{eq:pf-Txy}). The presence of stable eigenvalues in the last column implicates extra tangential stress term in introducing the predicted instabilities. }
    \label{table:physical mechanism}
  \end{center}
\end{table}

 \subsection{ Hadamard instability} \label{sec:Hadamard instability}

The dispersion curves shown in figure~\ref{fig:ci_vs_k_w1_B_vary} showed instability even at high wavenumber, i.e., Hadamard instability. In this section, we demonstrate that the Hadamard instability can be removed by adding stress diffusivity term in the \cite{Saramito-2007} constitutive equation~\eqref{eq:SM-equation}. Note that, the stress diffusion term has been previously utilised by \cite{fielding2005} to ascribe finite length to the shear bands in plane shear flow of \cite{johnson-segalman-1977} fluid.
Additionally, \cite{Buza-et-al-2022, Beneitez-et-al-2023, Couchman-et-al-2024} predicted the existence of a new polymer diffusive instability in a PCF of a viscoelastic fluid nonzero solvent contribution. Here, we do not consider solvent contribution. 

Upon addition of the stress diffusion term, the dimensionless constitutive equation becomes
\begin{eqnarray}
W \left[ \frac{\partial \boldsymbol{\tau}}{\partial t} + (\mathbf{v} \cdot{\nabla}) \boldsymbol{\tau}- (\nabla\mathbf{v})^T \cdot\boldsymbol{\tau}-\boldsymbol{\tau} \cdot (\nabla\mathbf{v}) \right] +  \left[\frac{|\boldsymbol{\tau}_d|-B}{|\boldsymbol{\tau}_d|} \right]  \boldsymbol{\tau}  = \boldsymbol{\dot \gamma} + \epsilon \nabla^2 \boldsymbol{\tau},~~~~~ \label{eq:SM-equation diffusivity}
\end{eqnarray}
where $\epsilon = D^*G^*/(\eta^* R^{*2})$ is the scaled stress diffusivity with $D^*$ as the dimensional stress diffusivity. The continuity and momentum equations given in \eqref{eq:linearised-PCF} remain applicable. The linearised constitutive equation stress components are
\begin{eqnarray}
\nonumber
W [ik(\bar v_x-c) \tilde \tau_{xx}+\tilde v_y D\bar \tau_{xx}-2ik \bar \tau_{xx} \tilde v_x-2 D\bar v_x \tilde \tau_{xy}-2 \bar \tau_{xy} D \tilde v_x]\\
\nonumber
+ {\cal F } [1+(t_4-t_2) \bar \tau_{xx}]\tilde \tau_{xx} + {\cal F }(t_3-t_1) \bar \tau_{xx}\tilde \tau_{xy} - {\cal F }(t_4-t_2) \bar \tau_{xx}\tilde \tau_{yy}\\
=2ik\tilde v_x + \epsilon (D^2 - k^2)\tilde \tau_{xx},\\
\nonumber
W [ik(\bar v_x-c) \tilde \tau_{xy}+\tilde v_y D\bar \tau_{xy} -ik \bar \tau_{xy} \tilde v_x-D\bar v_x \tilde \tau_{yy}-ik \bar \tau_{xx} \tilde v_y-\bar \tau_{xy} D\tilde v_y]\\
\nonumber
+ {\cal F } [1+(t_3-t_1) \bar \tau_{xy}]\tilde \tau_{xy} + {\cal F } \bar \tau_{xy}(t_4-t_2) (\tilde \tau_{xx}-\tilde \tau_{yy})\\
=  (ik\tilde v_y+D\tilde v_x)  + \epsilon (D^2 - k^2)\tilde \tau_{xy},\\
W [ik(\bar v_x-c) \tilde \tau_{yy}-2ik \bar \tau_{xy} \tilde v_y]+ {\cal F }\tilde \tau_{yy}=2D\tilde v_y  + \epsilon (D^2 - k^2)\tilde \tau_{yy},
\end{eqnarray}
where the parameters ${\cal F },t_1, t_2,t_3$ and $t_4$ remain unchanged. In addition to the boundary conditions specified in \eqref{bcs}, six more boundary conditions will be necessary due to the stress diffusion term. Following \cite{fielding2005,Buza-et-al-2022,Beneitez-et-al-2023,Couchman-et-al-2024}, we consider vanishing stress derivatives at $y=0,1$,
\begin{eqnarray}
    D \tilde \tau_{xx} =0; \quad  D \tilde \tau_{xy} =0; \quad D \tilde \tau_{yy} =0.
\end{eqnarray}

\begin{figure}
    \centering
    \begin{subfigure}[b]{0.55\textwidth}
        \centering
        \hspace* {-1.2cm}
        \includegraphics[width=\textwidth]{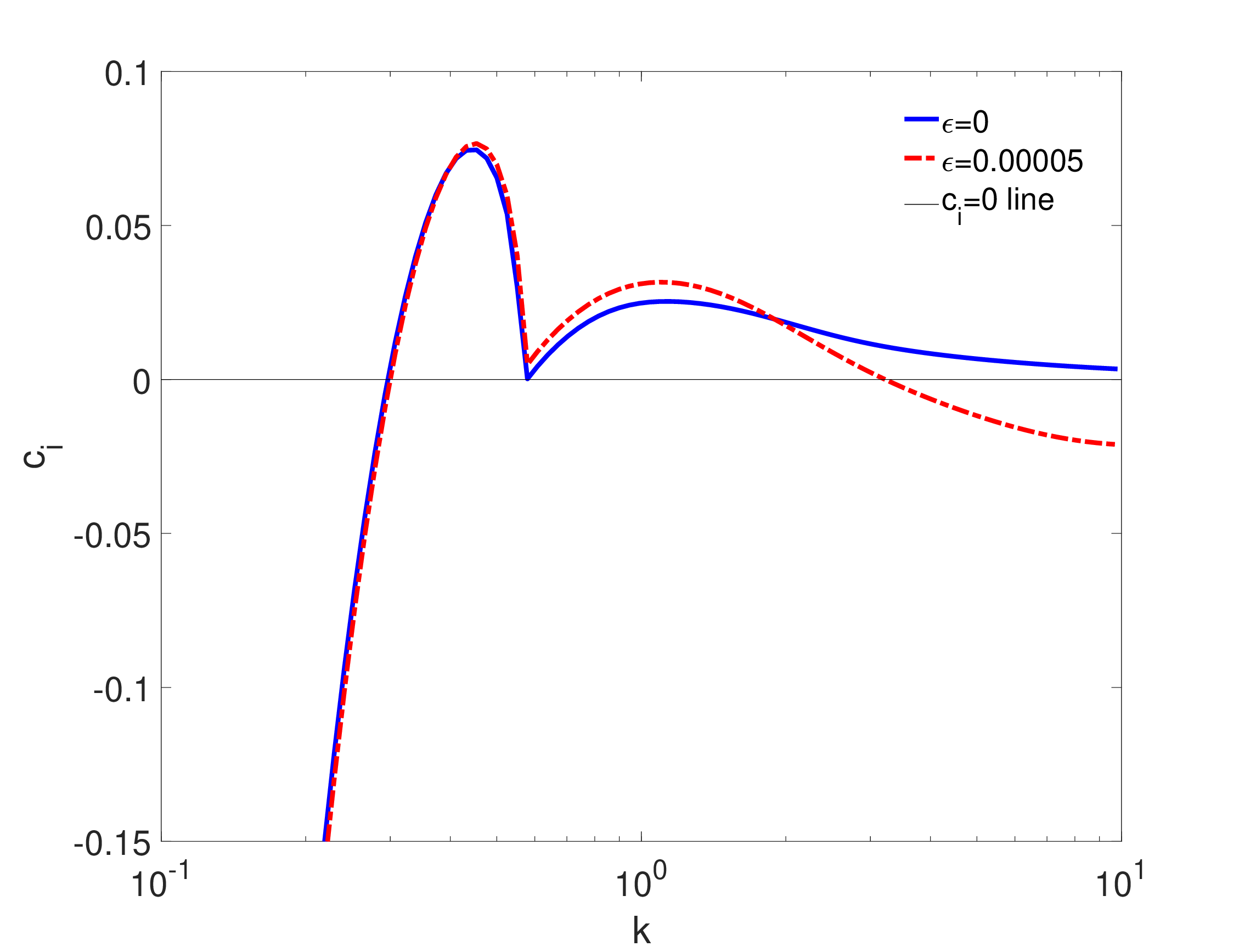}
        \caption{$W=1$ and $B=12$}
        \label{fig:ci_vs_k_w1_B12_d_vary}
    \end{subfigure}%
    ~
    \begin{subfigure}[b]{0.55\textwidth}
        \centering
        \hspace* {-1cm}
        \includegraphics[width=\textwidth]{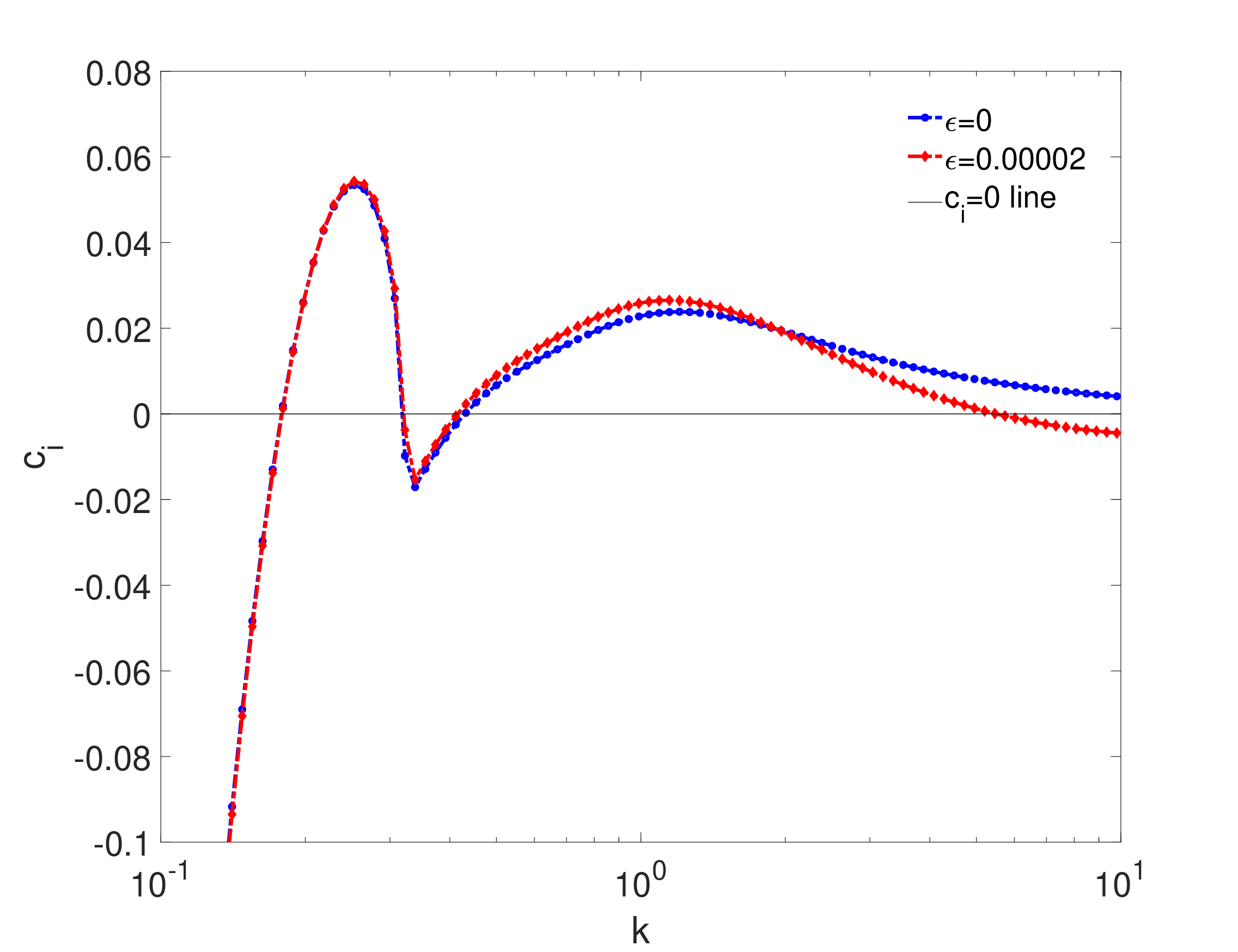} 
        \caption{$W=2$ and $B=20$}
        \label{fig:ci_vs_k_w2_B20_d_vary}
    \end{subfigure}
    \caption{ The effect of stress diffusivity on the dispersion curves. Panels show the stabilisation of the Hadamard instability as the stress diffusivity value increases for two sets of parameters.}
    \label{fig:dispersion plots hadamard instability}
\end{figure}

 The effect of nonzero stress diffusivity on the dispersion curves is shown in figure~\ref{fig:dispersion plots hadamard instability}. For $\epsilon=0$, the dispersion curve shows $c_i>0$ even for $k=10$. Although not shown here, this holds true for arbitrarily high wavenumber, signifying the presence of weak Hadamard instability. As $\epsilon$ assumes nonzero value even of $O(10^{-5})$, the perturbations at high wavenumber exhibit $c_i<0$, implying stabilisation of arbitrarily high wavenumber instability. Thus, the stress diffusivity succeeds in removing weak Hadamard instability.

\section{Conclusions} \label{sec:Conclusions}

The linear stability of an elastoviscoplastic (EVP) plane Couette flow (PCF) is analysed using \cite{Saramito-2007} model. The pseudo-spectral method based on Chebyshev polynomials is employed to solve the eigenvalue problem.  In the absence of yield stress, \cite{Saramito-2007} model reduces to the Upper Convected Maxwell (UCM) model. 

The PCF of UCM exhibits two stable GL modes in the creeping-flow limit. As the yield stress effect, i.e., Bingham number, is increased, the GL modes become unstable. Additionally, a new class of modes travelling at the average speed of the fluid are destabilised, termed here as \emph{centre modes}. Thus, analysis reveals two types of instabilities, viz., unstable GL modes and new centre modes. The analysis reveals that an extra tangential stress term arising due to the yield stress effect leads to the predicted instability. The dispersion curves are found to exhibit \emph{weak Hadamard instability}, i.e., unstable perturbations of arbitrarily high wavenumber. The analysis demonstrates that an addition of stress diffusion term in the \cite{Saramito-2007} model can remove the Hadamard instability.

\section*{Declaration of interest}

The author reports no conflict of interest.

 \section*{Acknowledgments}
The author acknowledges financial support from the SERB-DST under grant SRG/2023/000223.

\appendix

\bibliographystyle{jfm}
\bibliography{references}

\end{document}